# THERMODYNAMIC EFFICIENCY AND ENTROPY PRODUCTION IN THE CLIMATE SYSTEM

Valerio Lucarini [Email: valerio.lucarini@unibo.it]

Department of Physics, University of Bologna, Viale Berti Pichat 6/2, 40127 Bologna (Italy)

Istituto Nazionale di Fisica Nucleare – sezione di Bologna, Via Irnerio 46, 40127 Bologna (Italy)

CINFAI, Via Venanzi 15, 62032 Camerino (Italy)

**ABSTRACT**

We present a new outlook on the climate system thermodynamics. First, we construct an equivalent Carnot engine with efficiency $\eta$ and frame the Lorenz energy cycle in a macro-scale thermodynamic context. Then, by exploiting the 2$^{nd}$ law, we prove that the lower bound to the entropy production is $\eta$ times the integrated absolute value of the internal entropy fluctuations. An exergetic interpretation is also proposed. Finally, the controversial maximum entropy production principle is re-interpreted as requiring the joint optimization of heat transport and mechanical work production. These results provide new tools for climate change analysis and for climate models' validation.

## 1. INTRODUCTION

The analysis of the global thermodynamic properties of the climate system has long been the subject of an intense investigation, starting with the landmark analysis of the energy cycle of the atmosphere [1], which highlighted the concept of availability by showing that only a tiny part of the potential energy of the atmosphere can be converted to mechanical energy. Several authors have then addressed the issue of formalizing the concept of efficiency of the climate machine, driven by the temperature difference between a warm and a cold thermal bath. Often, those have been identified with the equatorial and polar region, which feature a positive and negative radiative balance at the top of the atmosphere, respectively. Thus, the atmospheric and oceanic motions can be interpreted both as the result of the mechanical work (then dissipated by viscosity) produced by the engine, and as tools able to re-equilibrate the energy balance of the climate system by turbulent heat transport [2,3]. Later on, in [4] it was introduced a formally more advanced analysis of the heat (and entropy) sources and sinks inside the climate system, thus allowing for a rigorous definition of a Carnot engine–equivalent picture of the climate machine.

After the publication of the landmark book [5], the thermodynamic analysis of non-equilibrium systems have gained more and more momentum, and is now widespread in engineering, chemistry, physics, biology, earth science, and many other fields. Non-equilibrium systems generate entropy by irreversible processes and keep a steady state by balancing the input and output of energy and entropy with the surrounding environment. Following the variational principle introduced in [6] for equilibrium statistical systems and, driven by the desire (and need) to find a guiding principle able to partially disentangle the complexity of non-equilibrium system, scholars of various disciples have conjectured the validity of the maximum entropy production principle (MEPP), which proposes that an out-of-equilibrium nonlinear system adjusts in such a way to maximize the production of entropy [7]. Note that, since a recent claim of a rigorous derivation of MEPP [8] has been rejected [9], the full understanding of the extent to which MEPP is valid and useful has not been attained.

A great deal of attention has been paid to the application of non-equilibrium thermodynamics to the climate system. Actually, some of the earlier stimulations towards the formulation of MEPP have come from the climate community [10]. In [3] a detailed theoretical presentation of the entropy production in the climate system and some reasonable estimates of its value are given, whereas in [11,12] a more modern perspective, which includes also applications of the MEPP, is provided. On a different note, the author, building upon the framework of non-equilibrium statistical mechanics [13] and of the response theory for non-equilibrium statistical systems [14], recently derived a set of universal constraints [15] useful for the analysis of climate-like systems [16].

In this paper we draw a line connecting the investigation of the climate as a thermal engine to the analysis of its entropy production, In Sect. 2, we revise the concept of efficiency [4] and present a more direct link to the energy cycle [1]. In Sect. 3, we exploit the 2$^{nd}$ law of thermodynamics to derive an inequality and relate an entropy production lower bound to the integrated absolute value of the entropy fluctuations of the system via the Carnot efficiency. This is then used to provide an interpretation of the MEPP and to motivate an exergetic analysis [17] of the system. In Sect. 4 we draw our conclusions.

## 2. THE CLIMATE SYSTEM AS A THERMAL ENGINE

Let the total energy of an $\Omega$-subdomain of the climatic system be [3]:

$$E(\Omega) = \int_\Omega dV \rho (u + \phi + k) \qquad (1)$$

where $\rho$ is the local density, $e = u + \phi + k$ is the total energy per unit mass, with $u$, $\phi$ and $k$ indicating the internal (inclusive of the contributions due to water phase transitions), potential and kinetic energy components, respectively. As the climate system is a multi-component one, the thermodynamic equations defining the medium are not the same everywhere (*e.g.* air vs. sea-water). The instantaneous balance of the energy of the system can be expressed as $\dot{E}(\Omega) = \dot{P}(\Omega) + \dot{K}(\Omega)$, where $P$ represents the integrated total potential energy (thermal + potential) and $K$ is the total kinetic energy. The time derivative of the total kinetic energy of the system is:

$$\dot{K}(\Omega) = -\int_\Omega dV \varepsilon^2 + C(P, K) = -\dot{D} + C(P, K) \qquad (2)$$

where the first term is the opposite of the integrated dissipation $\dot{D}$, with $\varepsilon^2$ being positive definite (by the 2$^{nd}$ law of thermodynamics), whereas the second term describes the net rate of conversion of potential into kinetic energy, as described in energy cycle formalism [1]. Therefore, we can interpret the second term as instantaneous work performed by the system and we denote it by $\dot{W} = C(P,K)$. When considering the total potential energy of the system, we have:

$$\dot{P}(\Omega) = \int_\Omega dV \rho \dot{Q} - \dot{W} \quad , \tag{3}$$

with $\dot{Q} = 1/\rho\left(\varepsilon^2 - \vec{\nabla}\cdot\vec{H}\right)$ representing the net hearting due to viscous processes and convergence of heat fluxes, which can be split into the radiative, sensible, and latent heat components. We obtain:

$$\dot{E}(\Omega) = \int_\Omega dV\left(-\vec{\nabla}\cdot\vec{H}\right) = -\int_{\partial\Omega} dS\hat{n}\cdot\vec{H} . \tag{4}$$

If the system is at steady state, for any subdomain $\Omega$, the quantities $E(\Omega)$, $P(\Omega)$, and $K(\Omega)$ are stationary (in terms of statistical properties). Therefore, $\overline{\dot{E}(\Omega)} = \overline{\dot{P}(\Omega)} = \overline{\dot{K}(\Omega)} = 0$, where the upper bar indicates time averaging over a long time scale. At any instant, we can partition the domain $\Omega$ into two subsets, $\Omega^+$ and $\Omega^-$, such that $Q(x) > 0, x \in \Omega^+$ and $Q(x) < 0, x \in \Omega^-$ [4]. Therefore:

$$\dot{P}(\Omega) + \dot{W} = \int_{\Omega^+} dV\rho\dot{Q}^+ + \int_{\Omega^-} dV\rho\dot{Q}^- = \dot{\Phi}^+ + \dot{\Phi}^- \tag{5}$$

where, by definition, $\dot{Q}^+$ and $\dot{Q}^-$ are positive and negative definite and the integrated quantities $\dot{\Phi}^+$ and $\dot{\Phi}^-$ are positive and negative at all times, respectively. Since dissipation is positive definite, we obtain $-\overline{\dot{K}(\Omega)} + \overline{\dot{W}} = \overline{\dot{D}} = \overline{\dot{P}(\Omega)} + \overline{\dot{W}} = \overline{\dot{W}} = \overline{\dot{\Phi}^+} + \overline{\dot{\Phi}^-} > 0$, with $\overline{\dot{\Phi}^+} > 0$ and $\overline{\dot{\Phi}^-} < 0$.

Using the second law of thermodynamics, we have $\dot{Q} \leq \dot{s}T$, where $\dot{s}$ is the time derivative of the entropy per unit mass. Assuming, as usual in climate dynamics, local thermodynamic equilibrium, the Clausius inequality reduces to $\dot{Q} = \dot{s}T$. Note that local thermodynamic equilibrium does not apply to the upper atmosphere, whose mass is, nevertheless, negligible. We hereby neglect the contribution to entropy variations due to mixing related to salinity fluxes in the ocean, and those due to the mixing of the water vapor in the atmosphere. In the first case, it has been estimated that such contribution is negligible as the entropy production related to salinity fluxes is three orders of magnitude smaller than what due to thermal processes [18]. As for the second case, we have that, locally, the magnitude of the contribution to entropy production due to water vapor mixing is $\approx |\rho(C-E)R_d \log(H)|$ [19], where $C$ and $E$ are instantaneous condensation and evaporation rates, $H$ is the relative humidity and $R_d$ is the dry air gas constant. Instead, the contribution due to the term representing the convergence of latent heat flux is $|\rho(C-E)L_v/T|$, where $L_v$ is the latent heat of vaporization. The ratio between the two terms can be estimated as $|\log(H)R_d T/L_v|$, which, using usual typical terrestrial values $H \approx 0.7$, $T \approx 250K$, results to be $|\log(H)R_d T/L_v| \approx 0.01$. Therefore, we feel that we can safely neglect the mixing processes in the rest of the discussion. Thus, the derivative $\dot{S}(\Omega)$ of the total entropy of the system can be approximated as:

$$\dot{S}(\Omega) = \int_\Omega dV \frac{\varepsilon^2 - \nabla \cdot \vec{H}}{T} = \int_{\Omega^+} dV \frac{\rho \dot{Q}^+}{T} + \int_{\Omega^-} dV \frac{\rho \dot{Q}^-}{T} = \int_{\Omega^+} dV \rho |\dot{s}^+| - \int_{\Omega^-} dV \rho |\dot{s}^-| = \dot{\Sigma}^+ + \dot{\Sigma}^- \quad (6)$$

where we have exploited the fact that $\dot{s}$ has always the same sign as $\dot{Q}$, so that at all times $\dot{\Sigma}^+ > 0$ and $\dot{\Sigma}^- < 0$. If we take long term average of the previous equation, since the system is at steady state, we have that $\overline{\dot{S}(\Omega)} = 0$, so that $\overline{\dot{\Sigma}^+} = -\overline{\dot{\Sigma}^-}$. This also implies that $2\overline{\dot{\Sigma}^+} = \overline{\int_\Omega dV \rho |\dot{s}|}$, so that $\overline{\dot{\Sigma}^+}$ measures the absolute value of the entropy fluctuations throughout the domain. Using the mean

value theorem, we obtain that $\overline{\dot{\Phi}^+} = \overline{\dot{\Sigma}^+}\Theta^+$ and $\overline{\dot{\Phi}^-} = \overline{\dot{\Sigma}^-}\Theta^-$, where $\Theta^+$ ($\Theta^-$) is the time and space averaged value of the temperature where absorption (release) of heat occurs. Since $\left|\overline{\dot{\Sigma}^+}\right| = \left|\overline{\dot{\Sigma}^-}\right|$ and $\left|\overline{\dot{\Phi}^+}\right| > \left|\overline{\dot{\Phi}^-}\right|$ we derive that $\Theta^+ > \Theta^-$, *i.e.* absorption typically occurs at higher temperature than release of heat [3,4]. By rearranging some of the formulas introduced in this section, we obtain that:

$$\overline{D} = \overline{W} = \overline{\dot{\Phi}^+} + \overline{\dot{\Phi}^-} = \overline{\dot{\Sigma}^+}\Theta^+ + \overline{\dot{\Sigma}^-}\Theta^- = \overline{\dot{\Sigma}^+}\left(\Theta^+ - \Theta^-\right) = \frac{\Theta^+ - \Theta^-}{\Theta^+}\overline{\dot{\Phi}^+} = \frac{\overline{\dot{\Phi}^+} + \overline{\dot{\Phi}^-}}{\overline{\dot{\Phi}^+}}\overline{\dot{\Phi}^+}. \tag{7}$$

The climate system can then be approximated as a Carnot engine whose warm and cold heat bath are at temperature $\Theta^+$ and $\Theta^-$, respectively. Therefore we obtain:

$$\overline{W} = \eta\overline{\dot{\Phi}^+} \tag{8}$$

where $\eta = \left(\Theta^+ - \Theta^-\right)/\Theta^+ = \left(\overline{\dot{\Phi}^+} + \overline{\dot{\Phi}^-}\right)/\overline{\dot{\Phi}^+}$ can rigorously be defined as the equivalent Carnot efficiency $\eta$ of the system. We need to remark that the consideration of long-term averages is not just a useful mathematical device, but rather provides the equivalent of ergodic averaging for the macro-system considered. As shown in [1] (and clarified in [4]), the long term average of the work performed by the system is equal to the long term average of the generation of available potential energy, which can be interpreted as the portion of the total potential energy which is available for reversible conversion. Note that this definition of efficiency is different from other ones proposed in the literature (see *e.g.* [11]) as it is related to the local heating and cooling processes occurring in the system. More commonly, efficiency is related to the ratio $\eta_{\overline{H_{in}}}$ between the long term averages of the work and of total energy flux $\overline{H_{in}}$ entering the system. We then have $\eta/\eta_{\overline{H_{in}}} = \overline{\dot{\Phi}^+}/\overline{H_{in}} < 1$.

## 3. ENTROPY PRODUCTION

The 2$^{nd}$ law of thermodynamics states that the entropy produced inside a system having temperature $T$ and receiving an amount of heat $\delta Q$ is larger than $\delta Q/T$ [20]. In our case we have:

$$\overline{\dot{S}_{in}(\Omega)} \geq \overline{\dot{S}_{min}(\Omega)} = \overline{\left(\frac{\int_\Omega dV \rho \dot{Q}}{\int_\Omega dV \rho T}\right)} = \overline{\left(\frac{\dot{\Phi}^+ + \dot{\Phi}^-}{\langle\Theta\rangle}\right)} \approx \frac{\overline{\dot{\Phi}^+ + \dot{\Phi}^-}}{\overline{\langle\Theta\rangle}} \approx \frac{\overline{\dot{\Phi}^+ + \dot{\Phi}^-}}{(\Theta^+ + \Theta^-)/2} = \frac{\overline{W}}{(\Theta^+ + \Theta^-)/2}, \quad (9)$$

where $\overline{\dot{S}_{in}(\Omega)}$ is the long-term average of the entropy production inside the system, $\overline{\dot{S}_{min}(\Omega)}$ is its minimal value, $\langle\Theta\rangle$ is the density-averaged temperature of the system. The approximation holds as long as we can neglect the impact of the time cross-correlation between the total net heat balance and the average temperature. Moreover, we assume that the density-averaged temperature can be approximated by the mean of the two *Carnot temperatures* $\Theta^+$ and $\Theta^-$. $\overline{\dot{S}_{min}(\Omega)}$ can thus be estimated as:

$$\overline{\dot{S}_{min}(\Omega)} \approx \frac{\overline{W}}{(\Theta^+ + \Theta^-)/2} = \frac{\eta}{(\Theta^+ + \Theta^-)/2}\overline{\dot{\Phi}^+} = \frac{\Theta^+ - \Theta^-}{(\Theta^+ + \Theta^-)/2}\frac{\overline{\dot{\Phi}^+}}{\Theta^+} = \frac{\Delta\Theta}{(\Theta^+ + \Theta^-)/2}\overline{\dot{\Sigma}^+}$$
$$= \eta\frac{\Theta^+}{(\Theta^+ + \Theta^-)/2}\overline{\dot{\Sigma}^+} \approx \eta\overline{\dot{\Sigma}^+} \quad (10)$$

where the last approximation holds as long as typical temperature differences are small with respect to the average temperature (as usual in the case of the climate system), or, operatively, if $\Delta\Theta/(\Theta^+ + \Theta^-) \ll 1$. Therefore, the thermodynamic efficiency of the system sets also the scale relating the minimal entropy production of the system – due to macroscopically irreversible processes - to the absolute value of the entropy fluctuations inside the system due to microscopically reversible heating or cooling processes. Note that if the system is isothermal and at equilibrium the internal entropy production is zero, since the efficiency $\eta$ is vanishing. This is agreement with the fact that the system has already attained its maximum entropy state.

In order to gain a better understanding of the entropy production of the fluid system, we need to frame jointly the entropy budget of the system and of its surroundings. The change of entropy of a system can be split into internally generated entropy plus the net entropy influx from the surrounding: $\Delta S(\Omega) = \Delta S_{in}(\Omega) + \Delta S_{ex}(\Omega)$ [5,11]. Going to instantaneous changes, we have:

$$\dot{S}(\Omega) = \dot{S}_{in}(\Omega) - \int_{\partial\Omega} dS\hat{n} \cdot \frac{\vec{H}}{T} \tag{11}$$

where local thermodynamic equilibrium is again assumed. Using Eq. (6) and considering that $\overline{\dot{S}(\Omega)} = 0$, we obtain that the long term average of the entropy production of the system is:

$$\overline{\dot{S}_{in}(\Omega)} = \int_{\partial\Omega} dS\hat{n} \cdot \overline{\left(\frac{\vec{H}}{T}\right)} = \int_{\Omega} dV \vec{\nabla} \cdot \overline{\left(\frac{\vec{H}}{T}\right)} = \int_{\Omega} dV \overline{\vec{H} \cdot \vec{\nabla}\left(\frac{1}{T}\right)} + \int_{\Omega} dV \overline{\left(\frac{\varepsilon^2}{T}\right)} \geq \overline{\dot{S}_{min}(\Omega)} \tag{12}$$

The two terms correspond to the entropy production due to down-gradient heat transport and to viscous dissipation, respectively. Since $\overline{\dot{D}} = \overline{\dot{W}}$, the second term can be approximated by $\overline{\dot{S}_{min}(\Omega)}$ as long as $\Delta\Theta/(\Theta^+ + \Theta^-) \ll 1$. Therefore, we have $\overline{\dot{S}_{in}(\Omega)} \approx \overline{\dot{S}_{min}(\Omega)}(1+\alpha) \approx \eta \overline{\Sigma^+}(1+\alpha)$, where:

$$\alpha = \int_{\Omega} dV \overline{\vec{H} \cdot \vec{\nabla}\left(\frac{1}{T}\right)} \bigg/ \overline{\dot{S}_{min}(\Omega)} \approx \int_{\Omega} dV \overline{\vec{H} \cdot \vec{\nabla}\left(\frac{1}{T}\right)} \bigg/ \int_{\Omega} dV \overline{\frac{\varepsilon^2}{T}} \tag{13}$$

is the ratio between the contributions to entropy production given by down-gradient heat transport and by viscous dissipation, respectively. Therefore, the more efficiently the system transports heat from high to low temperature regions, the larger is the entropy production, *ceteris paribus*. The *ceteris paribus* condition (or lack of) is crucial for interpreting several modeling studies on the climate system - see, e.g, [21,22] – showing that, by changing a diffusion-like parameter controlling the large scale heat transport, the entropy production is small for very strongly and very weakly diffusive systems, whereas the maximum is obtained for intermediate conditions. In Eqs. (12)-(13) we can see that, if heat transport down-gradient the temperature field is very strong, the efficiency $\eta$ is small because the difference between the temperatures of the warm and the cold reservoirs is greatly reduced (the system is almost isothermal), whereas, if the transport is very weak, the factor $\alpha$ is small. Therefore, the controversial MEPP - see *e.g.* [11,21] – cannot be naively interpreted as equivalent to the fact that the climate system, mostly through the instabilities of atmospheric and oceanic flow, tends to re-equilibrate energetically the equatorial and the polar regions [23]. In fact, MEPP requires a joint optimization of heat transport and of production of mechanical work.

A further characterization and quantification of the irreversibility of the climatic thermodynamical processes can be obtained by making use of the concept of exergy destruction (or anergy production), which is the decrease of energy available for conversion into mechanical work due to entropy-generating processes. This is a standard conceptual tool used in the analysis of engineered thermal system [17]. We can define the excess exergy destruction average rate $\overline{\Delta \dot{E}x_{des}(\Omega)}$ as that deriving from excess entropy production due to down-gradient heat transport, which is a process not leading to any mechanical energy production. We have:

$$\overline{\Delta \dot{E}x_{des}(\Omega)} \approx \overline{\langle\Theta\rangle}\left(\overline{\dot{S}_{in}(\Omega)} - \overline{\dot{S}_{\min}(\Omega)}\right) = \overline{\langle\Theta\rangle}\alpha\overline{\dot{S}_{\min}(\Omega)} \approx \alpha\eta\dot{\Phi}^{+} = \alpha\overline{\dot{W}(\Omega)}, \qquad (14)$$

so that, in this context, $\alpha$ results to be the ratio between the excess exergy destruction and mechanical energy generation average rates.

## 4. CONCLUSIONS

In this paper we have presented a succinct but thorough investigation of the global thermodynamical properties of the climate system, by analyzing the main implications of the 1$^{st}$ law and of the 2$^{nd}$ law of thermodynamics. Most of the results are, actually, of more general value, but the climate system provides especially outstanding stimulations and challenges. Following [4], we have first clarified the notion of efficiency by creating a formal analogue with an equivalent Carnot engine, and identifying the resulting mechanical work with the production (and eventual dissipation) of kinetic energy. Along these lines, it is possible to define, by suitable averaging procedures, two temperatures corresponding to the warm and cold heat reservoir, respectively, and to derive a Carnot-like expression for the efficiency of the climatic system. Such an approach provides a simple yet elegant thermodynamic macro-framework for the energy cycle [1].

We have then exploited the 2$^{nd}$ law of thermodynamics to determine a lower bound to the entropy production., which is approximately given by the Carnot efficiency times the absolute value of the internal entropy fluctuations of the system. We have then obtained that entropy production due to heat transport from hot to cold regions is basically the difference between the actual and the minimal entropy production. Since the more efficiently the system transports heat from high to low temperature regions, the larger is the entropy production, *ceteris paribus*, the controversial MEPP could naively be interpreted as optimality of climate system in the re-equilibration of the radiative imbalance between the equatorial and the polar regions. Instead, MEPP is shown to be roughly equivalent to the joint optimization of heat transport down-gradient the temperature field and of the production of mechanical work. This view of entropy production clarifies some results presented in [21,22], where it was shown that, by tuning the large scale heat transport, the entropy production is small for very strong and very weak diffusive systems, having respectively a negligible efficiency and a weak heat transport, whereas the maximum is obtained for intermediate conditions. Finally, an exergetic point of view, more typically adopted for the analysis of engineered thermal systems, is proposed, leading to the result that ratio between the entropy generation due to heat transport and the minimal entropy generation is the same as that between excess exergy destruction and mechanical energy generation average rates.

These results may provide useful concepts for the understanding of the global properties of a paradigmatic non-equilibrium statistical system as the climatic one, and may provide crucial benchmarks for the definition of metrics and diagnostic tool for the validation of climate models [24,25]. In fact, since the $2^{nd}$ law of thermodynamics is as fundamental as the $1^{st}$ law, it is proposed that the defined macro-thermodynamic parameters such as the thermodynamic efficiency, the equivalent Carnot temperatures, the entropy production, and exergy destruction of the system should be addressed as carefully as energy balance properties for defining the basic features of the climate system and of the outputs of climate models, as well as for providing rigorous measures of climate change. Note that the present results apply equally well for describing the thermodynamic properties of fluids enveloping general planetary systems. Ongoing and foreseen investigations include the actual calculation of the discussed thermodynamic parameters in simulations performed with climate models under a variety of conditions, as determined by the atmospheric composition, the land-sea mask, and the value of the astronomical parameters. Such an effort poses additional challenges, as commonly used numerical schemes are responsible for spurious entropy production [4,26], so that our approach might also be useful for devising strategies aimed at the improvement of the very structure of climate models.

**Acknowledgments**
The author wishes to thank K. Fraedrich and P. Stone for crucial stimulations and an anonymous reviewer for constructive criticism aimed at improving the paper.